\documentclass[aps,nofootinbib,noshowpacs,showkeys,preprintnumbers,amsmath,amssymb]{revtex4}

\usepackage{bm}
\usepackage{amsmath}
\usepackage{slashed}

\usepackage{graphicx}
\usepackage{appendix}

\allowdisplaybreaks[4]

\newcommand{\ud}{\mathrm{d}}
\begin{document}

\title{\textit{CP} violation in $B_s$ decays mediated by two nearly-degenerate Majorana neutrinos\footnote{Phys.Rev.D 105 (2022) 073001, \url{https://link.aps.org/doi/10.1103/PhysRevD.105.073001}}}

\author{Guangshuai Zhang$^1$}

\author{Bo-Qiang Ma$^{1,2,3}$}
\email{mabq@pku.edu.cn}

\affiliation{$^1$
School of Physics and State Key Laboratory of Nuclear Physics and
	Technology, Peking University, Beijing 100871,
	China\\
	$^2$Center for High Energy Physics, Peking University, Beijing 100871, China\\
$^3$Collaborative Innovation Center of Quantum Matter, Beijing, China}

\begin{abstract}
We study the \textit{CP} violation in the lepton-number-violating process $B_s^0 \rightarrow D_s^- \mu^+ \mu^+ \pi^-$ and its \textit{CP}-conjugate process $\bar{B_s^0} \rightarrow D_s^+ \mu^- \mu^- \pi^+$ that are induced by two GeV-scale Majorana neutrinos with nearly-degenerate masses. Our result shows that the size of the \textit{CP} violation could become considerable if the mass difference between the two Majorana neutrinos is around the decay width of the neutrinos. We perform experimental analysis on the \textit{CP}-violating processes at LHCb within its upgrade II. The analysis shows that under current constraint on the heavy-light neutrino mixing parameter, it is possible that such \textit{CP} violation is observed with the LHCb experimental ability. We also give the upper bound on the heavy-light mixing parameter $|U_{\mu N}|^2$ under the assumption that no positive signal of the process is observed.  The result shows that such modes can give complementary constraint compared with previous experiments such as NuTeV, BEBC, etc. in the mass region 1~GeV $< m_N <$ 3~GeV.
\end{abstract}

\maketitle

\section{Introduction}
The neutrino oscillation experiments have confirmed that at least two of the three generations of neutrinos have nonzero masses~\cite{Super-Kamiokande:1998kpq, Super-Kamiokande:2010orq}. The relation between the flavor eigenstates and mass eigenstates of three generations of neutrinos is described by the Pontecorvo-Maki-Nakagawa-Sakata (PMNS) matrix. Thus, two questions about the nature of the neutrino arise. The first is the origin of the neutrino mass. As a well-known approach for the generation of neutrino mass, the seesaw mechanism offers a natural explanation for the tininess of the neutrino mass, for which the introduction of one or more generations of heavy Majorana neutrinos is necessary~\cite{Minkowski:1977sc, GellMann:1980vs, Mohapatra:1979ia, Yanagida:1980xy}. Studies on the cosmological effect of the heavy neutrinos have shown that these heavy neutrinos can generate the observed baryon asymmetry of the Universe through leptogenesis~\cite{Canetti:2012kh}, and such heavy neutrinos can also serve as natural candidates for dark matter~\cite{Dodelson:1993je,Shi:1998km,Asaka:2005pn,Canetti:2012vf}.

Another question is whether a neutrino is a Majorana particle, i.e., its antiparticle is identical to itself, or not. If the neutrino is a Dirac particle, then the lepton number is conserved ($\Delta L = 0$). While if it is a Majorana particle, then the conservation of the lepton number can be violated by 2 units ($\Delta L = 2$). Thus, the most important approach to establish the Majorana nature of the neutrino is to search for the lepton-number-violating (LNV) processes. Up to now, the neutrinoless double beta ($0\nu \beta \beta$) decay is the most explored LNV channel~\cite{DellOro:2016tmg, Dolinski:2019nrj, Engel:2016xgb}. 
Another way is to look for LNV processes that are induced by Majorana neutrino in the hardron or $\tau$ lepton decays. 
In literature, the LNV processes in the decays of mesons ($K, D, D_s, B, B_c$)~\cite{Abad:1984gh, Littenberg:1991ek, Littenberg:2000fg, Ali:2001gsa, Ivanov:2004ch, Dib:2000wm, Atre:2005eb, Atre:2009rg,Helo:2010cw, Cvetic:2010rw, Cvetic:2016fbv, Cvetic:2017vwl, Cvetic:2019shl,  Milanes:2016rzr, Mejia-Guisao:2017gqp, Milanes:2018aku, Asaka:2016rwd,Chun:2019nwi, Quintero:2011yh}, baryons ($\Sigma^-, \Xi^-, \Lambda_b$)~\cite{Barbero:2002wm, Barbero:2007zm, Barbero:2013fc, Mejia-Guisao:2017nzx, Zhang:2021wjj}, and $\tau$ lepton~\cite{Castro:2012gi, Gribanov:2001vv, Cvetic:2002jy, Atre:2005eb} were extensively investigated. If the mass of the hypothetical Majorana neutrino lies in the range from hundreds of keV to several GeV, the decay widths of the corresponding LNV hadron/$\tau$ decays can be resonantly enhanced due to the on shellness of the intermediate Majorana neutrino. Thus, it is possible for these LNV decays to be observed by current or future hardron collider experiments. On the other hand, the nonobservation of these LNV hadron or $\tau$ decays can set strong constraints on the heavy-light neutrino mixing parameters in the resonant range. 

The neutrino oscillation experiments also showed that the $\theta_{13}$ angle in the PMNS matrix has a nonzero value~\cite{T2K:2011ypd, DayaBay:2012fng, RENO:2012mkc}; thus, the \textit{CP} violation in the neutrino sector is still possible. Moreover, the introduction of one or more generations of heavy neutrinos brings in more \textit{CP}-violating phases in the extended PMNS matrix, which describes the mixing between the flavor eigenstates and mass eigenstates of three normal neutrinos as well as the hypothetical sterile ones. Suppose there only exists one generation of sterile neutrino, the \textit{CP}-violating phases cause no observable effect in the sterile-neutrino-induced LNV processes.
However, if there exist two generations of GeV-scale sterile neutrinos that have nearly-degenerate masses, the \textit{CP}-violating phases in the extended-PMNS matrix could cause observable \textit{CP} violation in the sterile-neutrino-induced LNV meson decay. The idea was first pointed out in Ref.~\cite{Cvetic:2013eza}. The \textit{CP} violation in such Majorana-neutrino-induced LNV processes arises through two different mechanisms. The first is the interference between the amplitudes contributed by the two generations of sterile neutrinos. And the second is the neutrino oscillation during the propagation of the two on shell sterile neutrinos, which was first investigated in Ref.~\cite{Cvetic:2015ura} in LNV $B$ meson decays. \textit{CP}-violating LNV processes induced by similar mechanisms in the decays of other mesons~\cite{Cvetic:2014nla, Dib:2014pga, Cvetic:2015naa, Cvetic:2015ura, Cvetic:2020lyh, Zhang:2020hwj} and $\tau$ leptons~\cite{Zamora-Saa:2016ito, Tapia:2019coy} as well as $W$ boson~\cite{Najafi:2020dkp} were also studied in detail in the literature. Besides, the existence of two nearly-degenerate Majorana neutrinos together with the \textit{CP} violating phases can lead to a significant difference between the decay widths of the LNV meson decays and those of the lepton-number-conserving ones~\cite{Abada:2019bac, Das:2017hmg}, which is contrary to the common hypothesis. A remarkable conclusion about the \textit{CP} violation in LNV meson decays due to intermediate sterile neutrino interference is that the relative size of the \textit{CP} violation is independent of the neutrino mass while the parameter $\Delta m_N/\Gamma_N$ remains unchanged, where $\Delta m_N$ is the masses difference and $\Gamma_N$ the decay width of the neutrino. It is still unclear whether the conclusion holds true for four-body decays. Thus in this paper, we would apply the mechanism to four-body decays of $B_s$ mesons, of which the Feynman diagram is shown in Fig.~\ref{feynman diagram}. To our knowledge, no previous studies have explored the Majorana-neutrino-induced LNV decays of $B_s$ meson except Ref.~\cite{Mejia-Guisao:2017gqp}. Neither has any experimental research about the process been done. The mass difference between $B_s$ and $D_s$ reaches 3.4~GeV, which may extend the constraining mass region provided by previous channels such as three-body $D$ and $B$ meson decays. On the other hand, the branching fraction of $B_s \rightarrow D_s l \nu$ in $B_s$ decay reaches $8.1 \%$, which means the branching fraction of the related LNV processes may be considerable since the latter is proportional to the former in case that $m_N$ equals zero.

The motivation for the existence of two heavy neutrinos with nearly-degenerate masses comes from the $\nu$MSM model~\cite{Asaka:2005pn, Gorbunov:2007ak}, which proposes the existence of two generations of Majorana neutrinos with almost degenerate masses between 100~MeV and a few GeV as well as a light Majorana neutrino of mass $\sim$ 10~keV. The $\nu$MSM allows one to explain simultaneously neutrino oscillations, dark matter and baryon asymmetry of the Universe~\cite{Gorbunov:2007ak}.

In this work we study the \textit{CP} violation between the Majorana-neutrino-mediated LNV process $B_s^0 \rightarrow D_s^- \mu^+ \mu^+ \pi^-$ and its \textit{CP}-conjugate process $\bar{B_s^0} \rightarrow D_s^+ \mu^- \mu^- \pi^+$. These processes are induced by two Majorana neutrinos that have nearly degenerate but not equal masses. We focus on the neutrino mass region between 0.5 and 3.5~GeV where the resonant enhancement appears. We deal with the \textit{CP} violation that arises from the interference between the amplitudes contributed by the two Majorana neutrinos. Moreover, we investigate the possibility of the detection of \textit{CP} asymmetries in such decays during the LHCb upgrade II~\cite{LHCb:2018roe}.

The structure of this paper is as follows. In Sec.~\ref{section two}, we introduce the formalism we used for calculation and give the expression for the size of \textit{CP} violation. In Sec.~\ref{section three}, we perform an experimental analysis on the processes under the experimental ability of LHCb during its upgrade II and give the upper bounds on $|U_{\mu N}|^2$ under the assumption that such modes were not observed in experiments. Sec.~\ref{section four} gives the summary and the conclusion of our work.

\section{Formalism}
\label{section two}
\subsection{Decay widths for the two processes}

We define the light neutrino flavor eigenstates as
\begin{equation}
    \nu_l = \sum_{i=1}^{3} U_{l \nu_i}\nu_{i} + U_{l N_1}N_1 + U_{l N_2}N_2.
\end{equation}
Here, $\nu_i~(i=1,2,3)$ and $N_j~(j=1,2)$ represent the mass eigenstates of a light neutrino and heavy neutrino, respectively. $U_{l N_j}~(j=1,2)$ is the heavy-light neutrino mixing elements of the extended PMNS matrix (between $l$ lepton and the $j$th  heavy neutrino). We parametrize $U_{l N_i}$ as
\begin{equation}
    U_{l N_i} = |U_{l N_i}| e^{i\phi_{l N_i}}.
\label{CP phase}
\end{equation}
Here, $\phi_{l N_i}$ is the \textit{CP}-odd phase angle. We also require that the masses of $N_1$ and $N_2$ satisfy that
\begin{equation}
\begin{aligned}
    m_{\mu} + m_{\pi} \leq m_{N_1} &\leq m_{B_s} - m_{D_s} - m_{\mu},\\
    ~m_{\mu} + m_{\pi} \leq m_{N_2} &\leq m_{B_s} - m_{D_s} - m_{\mu},\\
\end{aligned}
\end{equation}
so that the intermediate neutrinos are nearly on shell and thus, the resonant enhancement appears. Another assumption we make is that the masses of the two neutrinos are nearly degenerate (suppose $m_{N_1} < m_{N_2}$),
\begin{equation}
    \Delta m_N = m_{N_2} - m_{N_1} \ll m_{N_i} ~(i=1,2).
\label{degenerate1}
\end{equation}

\begin{figure}[htb]
\includegraphics[width = 10cm]{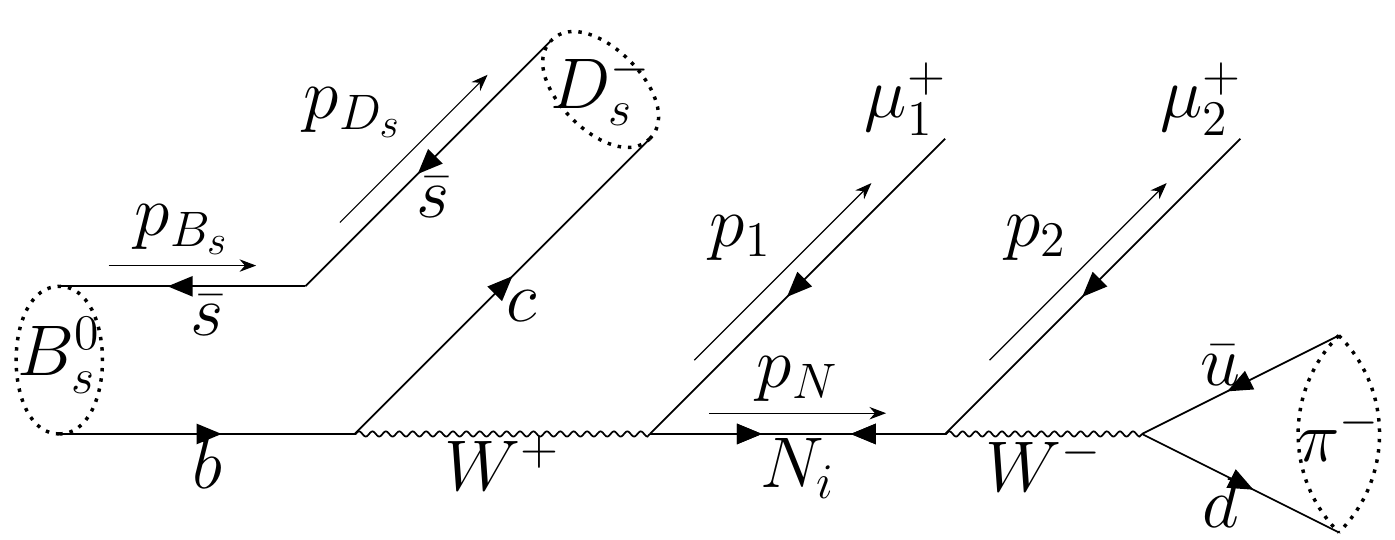}
\includegraphics[width = 10cm]{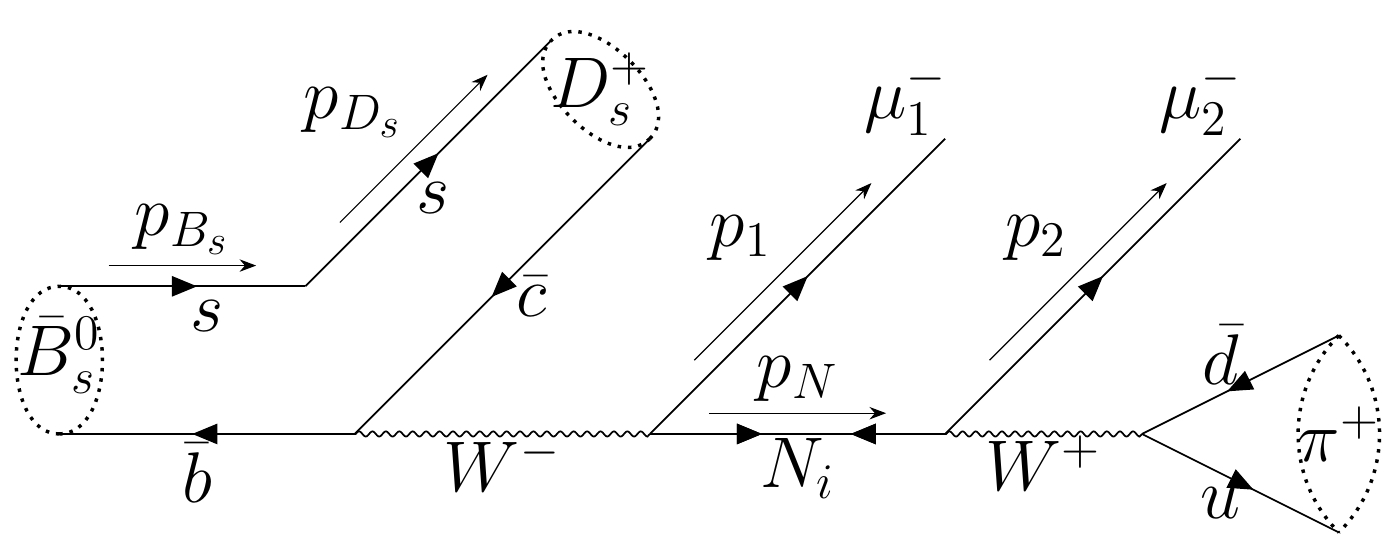} 
\caption{Feynman diagrams for $B_s^0 \rightarrow D_s^- \mu^+ \mu^+ \pi^-$ and $\bar{B}_s^0 \rightarrow D_s^+ \mu^- \mu^- \pi^+$}
\label{feynman diagram}
\end{figure}

The Feynman diagrams for the two processes are shown in Fig.~\ref{feynman diagram}.\footnote{There exists a ``crossed" diagram where $\mu_1$ and $\mu_2$ are exchanged. Since when the lifetime of the intermediate heavy neutrino is long enough, the two leptons appear at displaced vertices and the corresponding processes of ``direct" channel and ``cross" channel can be distinguished from each other~\cite{Dib:2014iga}. Thus, there is no interference between them and the corresponding decay width can be added directly. This brings in a factor 2 in the final result, which cancels out with the factor 1/2! due to the indistinguishability between the two muons .}. We denote the amplitude of $B_s^0 \rightarrow D_s^- \mu^+ \mu^+ \pi^-$ as $\mathcal{M}^+$ and that of its \textit{CP}-conjugate process as $\mathcal{M}^-$. The amplitudes can be written explicitly from the Feynman diagram
\begin{equation}
\begin{aligned}
    &i\mathcal{M}^+ = G_F^2 V_{cb}^*V_{ud}^*f_{\pi}^*\sum_{i = 1}^{2} U_{\mu N_i }^{*2}m_{N_i}P_{N_i}\bar{u}(\mu_2)\slashed{p}_{\pi}\gamma_{\mu}(1 - \gamma_5)v(\mu_1)\langle D_s^- |c(0) \gamma^{\mu} b(0)| B_s^0 \rangle,\\
    &i\mathcal{M}^- = G_F^2 V_{cb}V_{ud}f_{\pi}\sum_{i = 1}^{2} U_{\mu N_i }^{2}m_{N_i}P_{N_i}\bar{u}(\mu_2)\slashed{p}_{\pi}\gamma_{\nu}(1 + \gamma_5)v(\mu_1)\langle D_s^+ |\bar{c}(0) \gamma^{\nu} \bar{b}(0)| \bar{B}_s^0 \rangle.
\end{aligned}
\label{amplitude}
\end{equation}
Here, $G_F = 1.1664 \times 10^{-5} \mathrm{GeV}^{-2}$ is the Fermi coupling constant, and $f_\pi$ = 0.1304 GeV is the pion decay constant~\cite{Zyla:2020zbs}. $V_{cb}$ and $V_{ud}$ are the charm-bottom and upper-down Cabibbo-Kobayashi-Maskawa (CKM) matrix element. In this work we take that $|V_{ud}| = 0.974$ and $|V_{cb}| = 4.012\times 10^{-2}$~\cite{Zyla:2020zbs}. $m_{N_i}$ is the mass of the $i$th heavy neutrino. $\bar{u}(\mu_2)$ and $v(\mu_1)$ are the spinors of the two muons. The propagator $P_{N_i}$ is defined as
\begin{equation}
    P_{N_i} = \frac{1}{(p_{B_s} - p_{D_s} - p_{\mu_1})^2 - m_{N_i}^2 + i\Gamma_{N_i}M_{N_i}}.
\end{equation}
Also, $\langle D_s^- |c(0) \gamma_{\mu} b(0)| B_s^0 \rangle = \langle D_s^+ |\bar{c}(0) \gamma_{\mu} \bar{b}(0)| \bar{B}_s^0 \rangle$ is the $B_s-D_s$ transition matrix element that can be parametrized as~\cite{Monahan:2017uby}
\begin{equation}
    \langle D_s^- |c(0) \gamma_{\mu} b(0)| B_s^0 \rangle = f_0(t_1)\frac{m_{B_s}^2 - m_{D_s}^2}{t_1}q^{\mu} + f_+(t_1)\Big[p_{B_s}^{\mu} + p_{D_s}^{\mu} - \frac{m_{B_s}^2 - m_{D_s}^2}{t_1}q^{\mu}\Big],
\end{equation}
where $q^{\mu} = p_{B_s}^{\mu} - p_{D_s}^{\mu}$ is the transferred momentum and $t_1 = q^2$. In this work, we use the numerical result of the form factors $f_0(t_1)$ and $f_+(t_1)$ from Ref.~\cite{Monahan:2017uby} which is calculated by lattice QCD. 
Equation~(\ref{amplitude}) shows that two \textit{CP}-odd factors appear in the amplitudes that are nontrivial. The first is the \textit{CP}-odd phases of the heavy-light mixing parameter $U_{\mu N_i}$ and the second comes from the weak interaction vertex $\gamma^{\mu}(1 - \gamma_5)$. In case there exists only one generation of heavy neutrino, such a \textit{CP}-odd term would not result in a \textit{CP} violation in the observable quantity since after being squared, the difference between the two amplitudes vanishes. While in case there are more than one generation of heavy neutrino, a \textit{CP} violation would arise in the decay width. For simplicity, we write the amplitudes as
\begin{equation}
\begin{aligned}
    \mathcal{M^+} &= U_{\mu N_1}^{*2} \bar{\mathcal{M}}^+_1 + U_{\mu N_2}^{*2} \bar{\mathcal{M}}^+_2,\\
    \mathcal{M^-} &= U_{\mu N_1}^2 \bar{\mathcal{M}}^-_1 + U_{\mu N_2}^2 \bar{\mathcal{M}}^-_2.
\end{aligned}
\end{equation}
Here $\bar{\mathcal{M}}^{\pm}_i$ represents the canonical amplitude contributed by the $i$th heavy neutrino
\begin{equation}
\begin{aligned}
    \bar{\mathcal{M}}^+_i &= G_F^2 V_{cb}^*V_{ud}^*f_{\pi}^*m_{N_i}P_{N_i}\bar{u}(\mu_2)\slashed{p}_{\pi}\gamma_{\mu}(1 - \gamma_5)v(\mu_1)\langle D_s^- |c(0) \gamma^{\mu} b(0)| B_s^0 \rangle,\\
    \bar{\mathcal{M}}^-_i &= G_F^2 V_{cb}V_{ud}f_{\pi}m_{N_i}P_{N_i}\bar{u}(\mu_2)\slashed{p}_{\pi}\gamma_{\mu}(1 + \gamma_5)v(\mu_1)\langle D_s^- |c(0) \gamma^{\mu} b(0)| B_s^0 \rangle.
\end{aligned}
\end{equation}
It is straightforward to prove that
\begin{equation}
    \bar{\mathcal{M}}^{+}_i\bar{\mathcal{M}}^{+ *}_j = \bar{\mathcal{M}}^{-}_i\bar{\mathcal{M}}^{- *}_j.
\end{equation}
We then define the squared amplitude matrix $M$ as
\begin{equation}
    M_{ij} = \bar{\mathcal{M}}^{+}_i\bar{\mathcal{M}}^{+ *}_j = \bar{\mathcal{M}}^{-}_i\bar{\mathcal{M}}^{- *}_j.
\end{equation}
The explicit form of $M$ is shown in Appendix~\ref{appendix b}. From Equation~(\ref{M matrix}) in Appendix~\ref{appendix b} we can verify that
\begin{equation}
    M_{12} = M_{21}^*.
\end{equation}
Then the decay width for the two processes can be written as
\begin{equation}
\begin{aligned}
    \Gamma^+ &= \int d\Phi_4\Big[|U_{\mu N_1}|^4 M_{11} + |U_{\mu N_2}|^4 M_{22} + U_{\mu N_1}^{*2}U_{\mu N_2}^{2}M_{12} + U_{\mu N_1}^{2}U_{\mu N_2}^{*2}M_{21}\Big],\\
    \Gamma^- &= \int d\Phi_4\Big[|U_{\mu N_1}|^4 M_{11} + |U_{\mu N_2}|^4 M_{22} + U_{\mu N_1}^{2}U_{\mu N_2}^{*2}M_{12} + U_{\mu N_1}^{*2}U_{\mu N_2}^{2}M_{21}\Big],
\label{decay width}
\end{aligned}
\end{equation}
where $d\Phi_4$ is the four-body phase space
\begin{equation}
    d\Phi_4 = \frac{1}{2 m_{B_s}}\frac{1}{(2\pi)^8}\frac{d^3 \bm{p}_{D_s}}{(2 E_{D_s})}\frac{d^3 \bm{p}_{1}}{(2 E_{1})}\frac{d^3 \bm{p}_{2}}{(2 E_{2})}\frac{d^3 \bm{p}_{\pi}}{(2 E_{\pi})}\delta^4(p_{B_s} - p_{D_s} - p_{\mu_1} - p_{\mu_2} - p_{\pi}).
\end{equation}
The explicit form of $d\Phi_4$ and the reference frame we use to define the kinematic variables for numerical calculation are given in Appendix~\ref{appendix a}. Note that the factors $|P_{N_i}|^2$ appears in Equation~(\ref{M matrix}), which can be approximated as
\begin{equation}
    |P_{N_i}|^2 = |\frac{1}{p_N^2 - m_{N_i}^2 + i\Gamma_{N_i}m_{N_i}}|^2
    \approx \frac{\pi}{m_{N_i}\Gamma_{N_i}}\delta(p_N^2 - m_{N_i}^2)~(i=1,2),
\end{equation}
when it is satisfied that $\Gamma_{N_i} \ll m_{N_i}~ (i=1,2)$. Thus with the narrow width approximation, $\int d\Phi_4 M_{ii}$ can be simplified as
\begin{equation}
    \int d\Phi_4 M_{ii} = \bar{\Gamma}(B_s \rightarrow D_s \mu_1 N_i)\frac{\bar{\Gamma}(N_i \rightarrow \mu_2 \pi)}{\Gamma_{N_i}}.
\label{dphi4Mii}
\end{equation}
Here $\bar{\Gamma}(B_s \rightarrow D_s \mu_1 N_i)$ and $\bar{\Gamma}(N_i \rightarrow \mu_2 \pi)$ are the canonical decay widths for the subprocesses $B_s \rightarrow D_s \mu_1 N_i$ and $N_i \rightarrow \mu_2 \pi$, respectively,
\begin{equation}
    \bar{\Gamma}(B_s \rightarrow D_s \mu_1 N_i) = \frac{\Gamma(B_s \rightarrow D_s \mu_1 N_i)}{|U_{\mu N_i}|^2},\quad \bar{\Gamma}(N_i \rightarrow \mu_2 \pi) = \frac{\Gamma(N_i \rightarrow \mu_2 \pi)}{|U_{\mu N_i}|^2}.
\label{canonical_decay_width}
\end{equation}

The last part that is yet unknown is the decay width for the heavy neutrinos $\Gamma_{N_i}$. In the literature, $\Gamma_{N_i}$ can be calculated by summing up all possible decaying channels of the Majorana neutrino, which is explained in Appendix~\ref{appendix d}. From Equation~\ref{total decay width} as well as the nearly-degenerate condition $m_{N_1} \approx m_{N_2}$, we can simplify the ratio between $\Gamma_{N_1}$ and $\Gamma_{N_2}$ as
\begin{equation}
    \frac{\Gamma_{N_2}}{\Gamma_{N_1}} \approx \frac{|U_{e N_2}|^2 a_{e}(m_{N_1}) + |U_{\mu N_2}|^2 a_{\mu}(m_{N_1}) + |U_{\tau N_2}|^2 a_{\tau}(m_{N_1})}{|U_{e N_1}|^2 a_{e}(m_{N_1}) + |U_{\mu N_1}|^2 a_{\mu}(m_{N_1}) + |U_{\tau N_1}|^2 a_{\tau}(m_{N_1})},
\label{gammaN1/N2}
\end{equation}
where the meaning of the factors $a_{l}(l=e,\mu,\tau)$ is explained in Appendix~\ref{appendix d}. The equation will be used in the analytical analysis in the next subsection. However, here we emphasize that we treat the lifetime of the sterile neutrino $\tau_N = \hbar/\Gamma_N$ as a free parameter that can be measured by LHCb experiments in our experimental analysis. In Sec. III we explain in detail our treatment and justify that the treatment is consistent with the result in Appendix~\ref{appendix d}.

\subsection{\textit{CP} violation}
In the following part we define that
\begin{equation}
    \mathcal{A}_{CP} = \frac{S^-}{S^+}
\end{equation}
to measure the size of the \textit{CP} violation, where $S^{\pm}$ are defined as
\begin{equation}
    S^{\pm} = \Gamma^+ \pm \Gamma^-.
\end{equation}
Then, following Eqs~(\ref{decay width}) and (\ref{CP phase}), we have
\begin{equation}
    S^- = 4 |U_{\mu N_1}|^2|U_{\mu N_2}|^2 \sin(2\phi_{\mu N_1} - 2\phi_{\mu N_2}) \int d\Phi_4  \mathrm{Im}M_{12}.
\label{S^-}
\end{equation}
Here $\mathrm{Im}M_{12}$ represents the imaginary part of $M_{12}$. As shown by Appendix~\ref{appendix b}, the imaginary part of $M_{12}$ is proportional to the imaginary part of $P_{N_1}P_{N_2}^*$,
\begin{equation}
    \mathrm{Im}P_{N_1}P_{N_2}^* = \frac{m_{N_2}\Gamma_{N_2}(p_N^2 - m_{N_1}^2) - m_{N_1}\Gamma_{N_1}(p_N^2 - m_{N_2}^2)}{[(p_N^2 - m_{N_1}^2)^2 + (m_{N_1}\Gamma_{N_1})^2][(p_N^2 - m_{N_2}^2)^2 + (m_{N_2}\Gamma_{N_2})^2]}.
\label{Impn}
\end{equation}
The physics meaning of Equation~(\ref{S^-}) includes two parts. First, the appearance of the factor $\sin (2\phi_{\mu N_1} - 2\phi_{\mu N_2})$ implies that the physics origin of the \textit{CP} violation is the difference between the \textit{CP}-odd phases of mixing parameters between the two heavy neutrinos with the common ones. We define the \textit{CP}-odd phase difference $\vartheta_{12}$ as
\begin{equation}
\vartheta_{12} = 2(\phi_{\mu N_1} - \phi_{\mu N_2}),    
\end{equation}
which is the key parameter in deciding on the size of the \textit{CP} violation. Second, the \textit{CP} violation of the processes in consideration arises as a result of the interference of the contributions of the two generations of Majorana neutrinos, which is shown explicitly by the factor $\mathrm{Im}P_{N_1}P_{N_2}^*$. 
In case that $m_{N_1} = m_{N_2}$, $\mathrm{Im}P_{N_1}P_{N_2}^*$ vanishes and so does the \textit{CP}-violating term $S^-$.

The sum $S^+$ can be written as
\begin{equation}
    S^+ = 2\int d\Phi_4 \Big[ |U_{\mu N_1}|^4 M_{11} + |U_{\mu N_2}|^4 M_{22} + 2|U_{\mu N_1}|^2|U_{\mu N_2}|^2\mathrm{Re}M_{12}\cos\vartheta_{12}\Big],
\end{equation}
where $\mathrm{Re}M_{12}$ represents the real part of $M_{12}$, of which the corresponding key factor is the real part of $P_{N_1}P_{N_2}^*$,
\begin{equation}
    \mathrm{Re}P_{N_1}P_{N_2}^* = \frac{(p_N^2 - m_{N_1}^2)(p_N^2 - m_{N_2}^2) + m_{N_1}m_{N_2}\Gamma_{N_1}\Gamma_{N_2}}{[(p_N^2 - m_{N_1}^2)^2 + (m_{N_1}\Gamma_{N_1})^2][(p_N^2 - m_{N_2}^2)^2 + (m_{N_2}\Gamma_{N_2})^2]}.
\end{equation} 
Note that Eqs~(\ref{dphi4Mii}) and (\ref{canonical_decay_width}) show that $\Gamma_{N_1}\times\int d\Phi_4 M_{11}$($\Gamma_{N_2}\times\int d\Phi_4 M_{22}$) is only the function of the heavy neutrino mass $m_{N_1}$($m_{N_2}$) and the two functions are of the same form. Considering the nearly-degenerate situation $m_{N_1} \approx m_{N_2}$, we have
\begin{equation}
    \frac{\int d\Phi_4 M_{22}}{\int d\Phi_4 M_{11}} \approx \frac{\Gamma_{N_1}}{\Gamma_{N_2}}.
\end{equation}
Then, the relative size of the \textit{CP} violation $\mathcal{A}_{CP}$ can be written in the following simple form
\begin{equation}
    \mathcal{A}_{CP} = \frac{2 \delta_{I} \kappa \sin\vartheta_{12}}{1 + \kappa^2 \frac{\Gamma_{N_1}}{\Gamma_{N_2}} + 2\kappa \delta_{R}\cos \vartheta_{12}}.
\end{equation}
Here, we define that
\begin{equation}
    \delta_{R} = \frac{\int d\Phi_4\mathrm{Re}M_{12}}{\int d\Phi_4 M_{11}},~
    \delta_{I} = \frac{\int d\Phi_4\mathrm{Im}M_{12}}{\int d\Phi_4 M_{11}},
\end{equation}
to measure the relative size of the interference term of the two heavy neutrinos. The parameter $\kappa$ is defined as
\begin{equation}
    \kappa = \frac{|U_{\mu N_2}|^2}{|U_{\mu N_1}|^2}.
\label{kappa}
\end{equation}

\begin{figure}[htb]
\includegraphics[width = 15 cm]{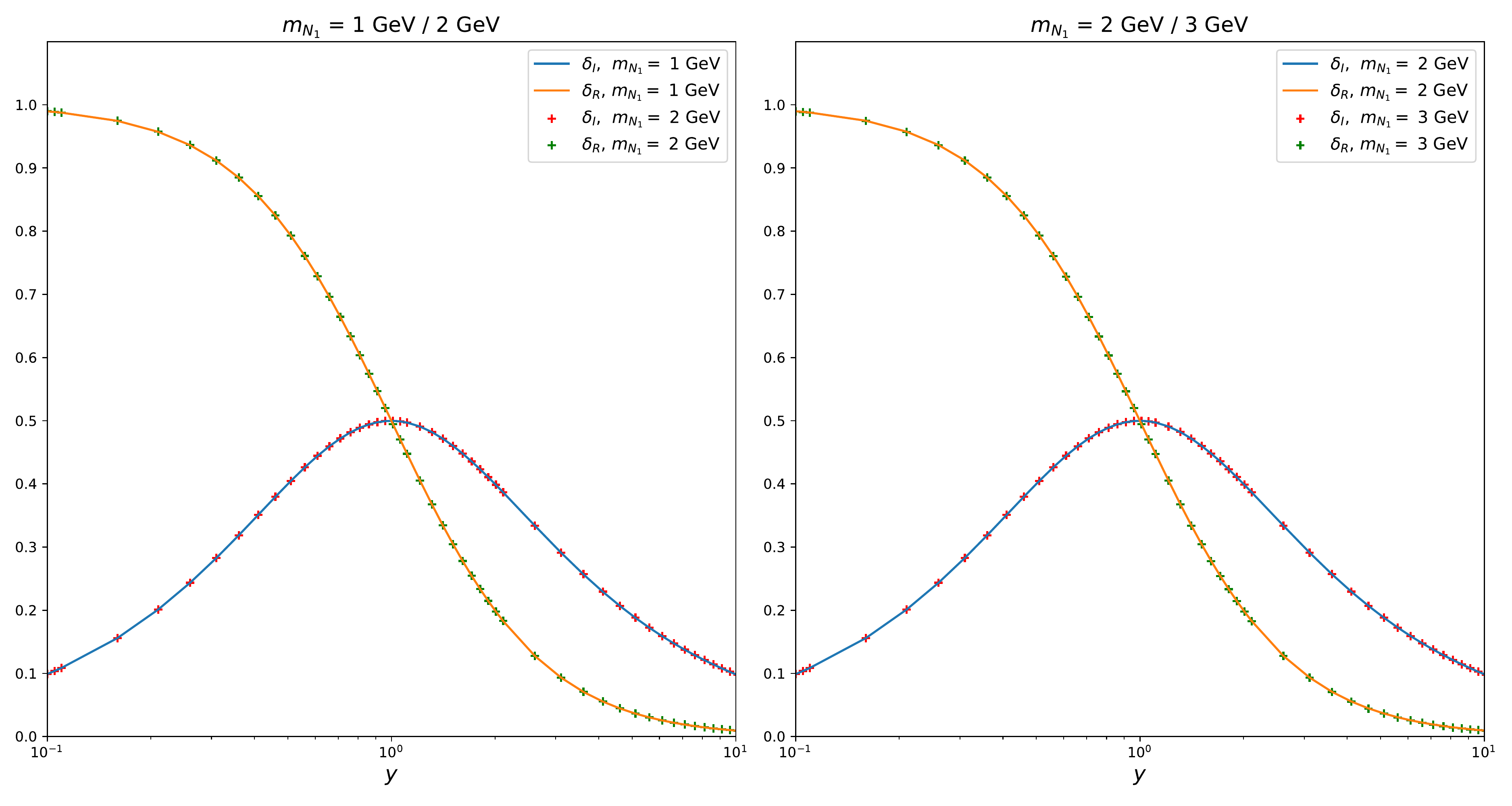}
\caption{The parameters $\delta_I$ and $\delta_R$ as a function of $y$ for different values of $m_{N_1}$. Note that since the deviation of $\delta_I/\delta_R$ between different choices of $m_{N_1}$ is very small (detailed numerical result shows that the difference is less than 1\%), to distinguish them, we use different styles of plot, i.e., line plot and scatter point plot, to represent the results for different choices of $m_{N_1}$.}
\label{delta1 and delta2}
\end{figure}

\begin{figure}[htb]
\includegraphics[width = 8cm]{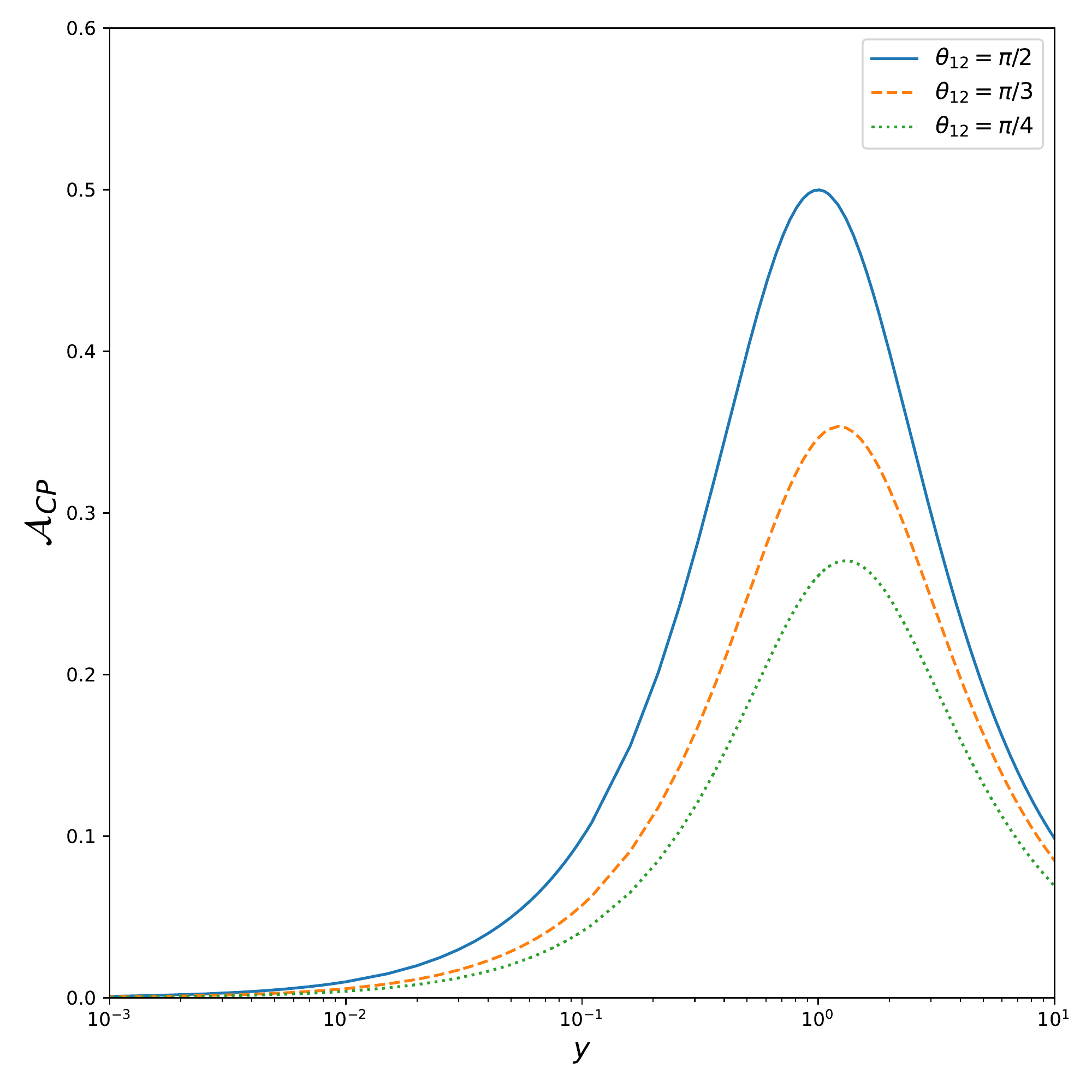}
\caption{The size of the \textit{CP} violation $\mathcal{A}_CP$ as a function of $y$ for different choices of $\vartheta_{12}$.}
\label{A_CP_plot}
\end{figure}

In order to make a more detailed analysis of this result, we make another assumption about the properties of the heavy neutrinos that the two Majorana neutrinos have approximately the same mixing parameters with the three generations of normal neutrinos, namely,
\begin{equation}
    \frac{|U_{l N_2}|^2}{|U_{l N_1}|^2} \approx 1 ~ (l=e, \mu, \tau).
\end{equation}
Thus, the right-hand side of Equation~(\ref{gammaN1/N2}) simply gives one. The size of the \textit{CP} violation $\mathcal{A}_{CP}$ can be further simplified as
\begin{equation}
    \mathcal{A}_{CP} = \frac{\delta_{I}\sin \vartheta_{12}}{1 + \delta_{R}\cos \vartheta_{12}}.
\label{A_CP}
\end{equation}
The final result Equation~(\ref{A_CP}) is only the function of $m_{N_1}$, $m_{N_2}$ and the angle $\vartheta_{12}$, or equivalently, $m_{N_1}$, $\Delta m_N$ and $\vartheta_{12}$. A natural way to measure the size of $\Delta m_N$ is to compare it with the decay width of the heavy neutrino $\Gamma_N$; thus, we define that
\begin{equation}
    y = \frac{\Delta m_N}{\Gamma_N}.
\end{equation}
Note that $\delta_I$ and $\delta_R$ rely both on the values of $m_{N_1}$ and $y$. However, numerical result shows that in the mass region under consideration, namely, 0.5~GeV $<m_{N_1}<$ 3.5~GeV, the variation of $\delta_I$ and $\delta_R$ due to different choices of $m_N$ is very small (less than $1\%$). In Fig.~\ref{delta1 and delta2}, we present the numerical results of $\delta_I$ and $\delta_R$ as a function of $y$ under the condition that $m_{N_1}$ equals 1, 2 and 3~GeV. In conclusion, $\delta_I$ and $\delta_R$ can be considered as nearly independent of the choice of $m_{N_1}$ in the mass region 0.5~GeV $<m_{N_1}<$ 3.5~GeV. Thus, the size of the \textit{CP} violation $\mathcal{A}_{CP}$ is also independent of $m_{N_1}$. The relative size of $\mathcal{A}_{CP}$ reaches its maximum when $y$ is around 1. The above conclusions are all in accordance with the three-body case studied in Refs.~\cite{Cvetic:2013eza, Cvetic:2015ura}.
Note that the conclusion does not hold true when the mass interval is larger than several tens of GeV. Reference~\cite{Najafi:2020dkp} studies \textit{CP} violation of a similar mechanism in $W$ boson decays and it shows that under the condition that $m_{N_1}$ equals 5, 10, 20 and 60~GeV, $\mathcal{A}_{CP}$ shows considerable differences. Another significant difference is that in our result $\mathcal{A}_{CP}$ reaches maximum when $y$ is around 1 while in Ref.~\cite{Najafi:2020dkp} $y$ is between 0.3 and 0.5. See Ref.~\cite{Najafi:2020dkp} for details. In Fig.~\ref{A_CP_plot}, we give the numerical result of $\mathcal{A}_{CP}$ as a function of $y$ under different choices of $\vartheta_{12}$.

Finally, the decay width for the process $B_s^0 \rightarrow D_s^- \mu^+ \mu^+ \pi^-$ and its \textit{CP}-conjugate process can be written in the more compact form,
\begin{equation}
    \Gamma^{\pm} = 2|U_{\mu N_1}|^4 [1 + \delta_{R}\cos \vartheta_{12} \pm \delta_{I}\sin\vartheta_{12}] \int d\Phi^4 M_{11},
\end{equation}
We define the averaged branching ratio $\mathcal{B}r_{avr}$ for experimental analysis in the next section,
\begin{equation}
    \mathcal{B}r_{avr} = \frac{1}{2\Gamma_{B_s}}(\Gamma^+ + \Gamma^-),
\end{equation}
here, $\Gamma_{B_s} = 4.362\times 10^{-13}~\mathrm{GeV}$ is the total decay width of $B_s$ meson~\cite{Zyla:2020zbs}.

\section{Experimental analysis}
\label{section three}

In this part, we evaluate the possibility that such \textit{CP} violation is observed by LHCb in its upgrade II~\cite{LHCb:2018roe}. In the experiment, the most important quantity is the absolute size of the averaged branching ratio $\mathcal{B}r_{avr}$. In this section, we do not distinguish the two sterile neutrinos due to the degeneracy and use $N$ to represent both of them.

The expected number of $B_s^0$ mesons produced on LHCb can be estimated as
\begin{equation}
    N_{B_s} = \mathcal{L} \times \sigma_{b\bar{b}} \times f(b \rightarrow B_s^0),
\end{equation}
where $\mathcal{L} \approx 200 ~ \mathrm{fb^{-1}}$ is the expected integrated luminosity of LHCb until 2035~\cite{LHCb:2018roe}, $\sigma_{b\bar{b}} \simeq 144~ \mathrm{\mu b}$ the $b\bar{b}$ cross section within the LHCb covered $\eta$ range ($2 < \eta < 5$)~\cite{Aaij:2016avz}, and $f(b \rightarrow B_s^0) \simeq 4.4 \% $ the hardronization factor of b quark to $B_s$ which is estimated from Ref.~\cite{Aaij:2019pqz}\footnote{Ref.~\cite{Aaij:2019pqz} shows that the ratio between the production fraction of $\Lambda_b$ hardrons and the sum of the fraction of $B^-$ and $\bar{B}^0$ is around 0.259 (averaged between $4~\mathrm{GeV} < p_T < 25~\mathrm{GeV}$ and $2 < \eta < 5$), while that between $\bar{B}^0_s$ and the sum of $B^-$ and $\bar{B}^0$ is 0.122. Since $B^{\pm}$, $B^0/\bar{B}^0$, $\bar{B}^{0}_s/B^0_s$ and $\Lambda_b^{0}/\bar{\Lambda}^0_b$ make the majority of $b\bar{b}$ products, the fraction $f(b\bar{b} \rightarrow \Lambda_b)$ is estimated as $0.122/(1+0.259+0.122)\times0.5 \sim 0.044$.}. The final result is $N_{B_s} \approx 1.9\times 10^{12}$.

In order to get the expected sensitivity of LHCb on the processes in consideration, we also need to know the detection efficiency of the LHCb detector $\epsilon~(B_s \rightarrow D_s \mu \mu \pi)$, which contains the contribution from geometrical acceptance, trigger and selection requirements and particle identification~\cite{LHCb:2013svv}. A systematic evaluation of the detection efficiency requires a Monte Carlo simulation under a LHCb configuration as well as considering final state radiation generation and interaction of the produced particles with the detector and its response~\cite{LHCb:2013svv}. Here, however, we use an indirect approach to give an approximation to the detection efficiency, whose accuracy is enough for our calculation. Reference~\cite{LHCb:2015wdu} shows that the simulated detection efficiency of the process $B_s^0 \rightarrow \phi(K^+K^-)\mu^+\mu^-$ is $1.1\%$. The main difference between $B_s^0 \rightarrow \phi(K^+K^-)\mu^+\mu^-$ and $B_s^0 \rightarrow D_s^- \pi^- \mu^+\mu^+$ comes from the replacement of $K^-$ with $D_s^-$ \footnote{The difference between the detection of $K$ and $\pi$ as well as between $\mu^+$ and $\mu^-$ is very small and can be overlooked here.}. $D_s$ is reconstructed by the golden mode $D_s \rightarrow K K \pi$, which requires two additional charged tracks and thus would reduce the detection efficiency. From Ref.~\cite{LHCb:2020ist}, we know that the ratio between the detection efficiency of $B_s^0 \rightarrow K^- \mu^+ \nu$ and $B_s^0 \rightarrow D_s^- \mu^+ \nu$ is around 0.733 (averaged over the full $q^2$ range). Thus, we estimate the detection efficiency of $B_s \rightarrow D_s \pi \mu \mu$ to be $1.1\% \times 0.733 \approx 0.81\%$. We note that the accuracy of this approximation is enough for magnitude estimation. 

Another factor we need to consider is the efficiency loss due to the flight of the long-lived particle, i.e., the intermediate sterile neutrinos $N$, in the detector. The sterile neutrinos are produced nearly on shell and would travel for certain distance before decaying into its aftermath~\cite{Dib:2014iga}. We include this effect by adding another factor $\mathcal{P}_N$ to the total detection efficiency, which relies on the lifetime of the sterile neutrino $\tau_N$ as
\begin{equation}
    \mathcal{P}_N = 1 - \exp{(-L_{D}/L_{N})},
\end{equation}
where $L_D \sim$ 1~m is the length of the detector and
\begin{equation}
   L_N = c \tau_N \gamma_N \beta_N 
\end{equation}
is the decay length of the sterile neutrino~\cite{Dib:2014iga}. Here $c$ represents the light speed and $\gamma_N \beta_N$ is the Lorentz time dilation factor of $N$. To our knowledge, the choice of $\gamma_N \beta_N$ between 1 and 10 is common in the literature such as Refs.~\cite{Cvetic:2016fbv, Dib:2014iga} based on the realistic condition of collider experiments. In this study, we take that $\gamma_N \beta_N$ equals 4 instead of doing detailed calculation of $\gamma_N \beta_N$ since the former is enough for magnitude estimation. For $\tau_N = 1000$~ps , the factor $\mathcal{P}_N$ is about $56\%$, while for $\tau_N \leq 100$~ps $\mathcal{P}_N$ is very close to 1 and the effect is almost negligible.

\begin{figure}[htb]
\includegraphics[width = 8cm]{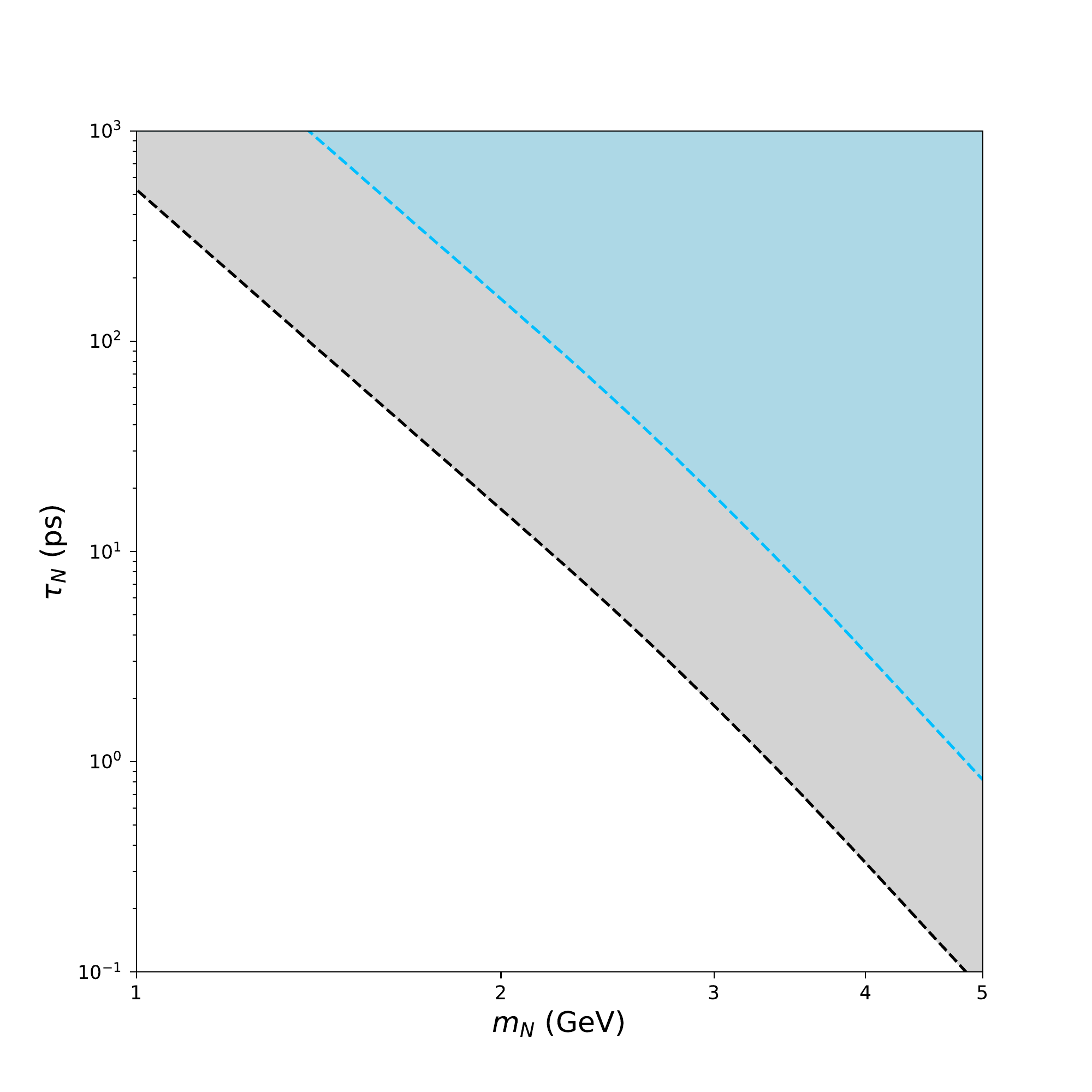}
\caption{The shaded region represent the possible region of the lifetime of the sterile neutrino $N$. The black line represents the case when $|U_{e N}|^2 = 10^{-7}, |U_{\mu N}|^2 = 10^{-4}, |U_{\tau N}|^2 = 10^{-2}$, and the blue line represents the case when $|U_{e N}|^2 = 10^{-8}, |U_{\mu N}|^2 = 10^{-5}, |U_{\tau N}|^2 = 10^{-3}$.}
\label{tau_N_plot}
\end{figure}

Following the practice in experimental research of sterile neutrinos such as Refs.~\cite{LHCb:2014osd, LHCb:2016inz}, we take $\tau_{N}$ as a free parameter that can be measured by LHCb experiment to avoid more complication, instead of calculating $\tau_N$ through $\tau_N = \hbar/\Gamma_N$. We assume that the sterile neutrino lifetime $\tau_N = [100, 1000]$~ps, which is within the acceptance of LHCb. Here, we justify that this assumption is consistent with Equation~(\ref{total decay width}) under a certain choice of the size of $|U_{lN}|^2$. It can be read from Ref.~\cite{Deppisch:2015qwa} that around 1 GeV the currently known upper bound on $|U_{e N}|^2$ is between $10^{-7}$ and $10^{-8}$ , and the upper bound on $|U_{\mu N}|^2$ is between $10^{-4}$ and $10^{-5}$, while that on $|U_{\tau N}|^2$ is between $10^{-3}$ and $10^{-2}$.  The possible region that the corresponding lifetime of the sterile neutrino can lie in is drawn in Fig.~\ref{tau_N_plot}. The plot shows that within the mass range $[1~\mathrm{GeV}, ~4~\mathrm{GeV}]$, the choice that $\tau_N = [100, 1000]$~ps is acceptable.

In Table.~\ref{event number}, we present the expected number of events for certain choices of related parameters, based on the experimental ability discussed above. We use $\mathcal{N}_+$/$\mathcal{N}_-$ to represent the event numbers of $B_s^0 \rightarrow D_s^- \mu^+ \mu^+ \pi^-$/$\bar{B_s^0} \rightarrow D_s^+ \mu^- \mu^- \pi^+$. The expected number of events at LHCb upgrade II is estimated as
\begin{equation}
    \mathcal{N}_{\pm} = N_{B_s}\times \epsilon~(B_s \rightarrow D_s \mu \mu \pi) \times \mathcal{P}_N \times \frac{\Gamma^{\pm}}{\Gamma_{B_s}}.
\end{equation}
For $|U_{\mu N}|^2 = 10^{-4}$, several hundreds of events can be expected, and there is significant difference between $\mathcal{N}_+$ and $\mathcal{N}_-$. For $|U_{\mu N}|^2 = 10^{-5}$, event numbers decrease to a few, and the observable \textit{CP} violation is not that significant. However, it should be noted that our result is based on a relatively conservative estimation on the experimental ability of LHCb since we use the detection efficiency of previous LHCb experiments. During the upgrade II, certain detection ability of LHCb detectors will be improved~\cite{LHCb:2018roe} and it is possible that more events can be observed.

\begin{table}[htb]
\centering
\begin{tabular}{ |p{2.5cm}|p{1.2cm}|p{1.2cm}|p{1.2cm}|p{1.2cm}| } 
  \hline
   &\multicolumn{2}{|c|}{$\tau_N = $ 1000~ps}&\multicolumn{2}{|c|}{$\tau_N = $ 100~ps}\\
  \hline
  & $\mathcal{N}_+$ & $\mathcal{N}_-$ & $\mathcal{N}_+$ & $\mathcal{N}_-$\\
  \hline
 $|U_{\mu N}|^2 = 10^{-4}$ & 665 & 390 & 119 & 70 \\
 \hline
 $|U_{\mu N}|^2 = 10^{-5}$ & 7 & 4 & 1 & 0 \\ 
 \hline
\end{tabular}
\caption{Expected event numbers for $B_s^0 \rightarrow D_s^- \mu^+ \mu^+ \pi^-$ ($\mathcal{N}_+$) and $\bar{B_s^0} \rightarrow D_s^+ \mu^- \mu^- \pi^+$ ($\mathcal{N}_+$) under the assumption that $|U_{\mu N}|^2$ equals $10^{-4}$ and $10^{-5}$. The relative parameters are chosen as follows: $m_N = 2~ \mathrm{GeV}$, $y = 1$, $\vartheta_{12} = \pi/4$.}
\label{event number}
\end{table}

\begin{figure}[htb]
\includegraphics[width = 12cm]{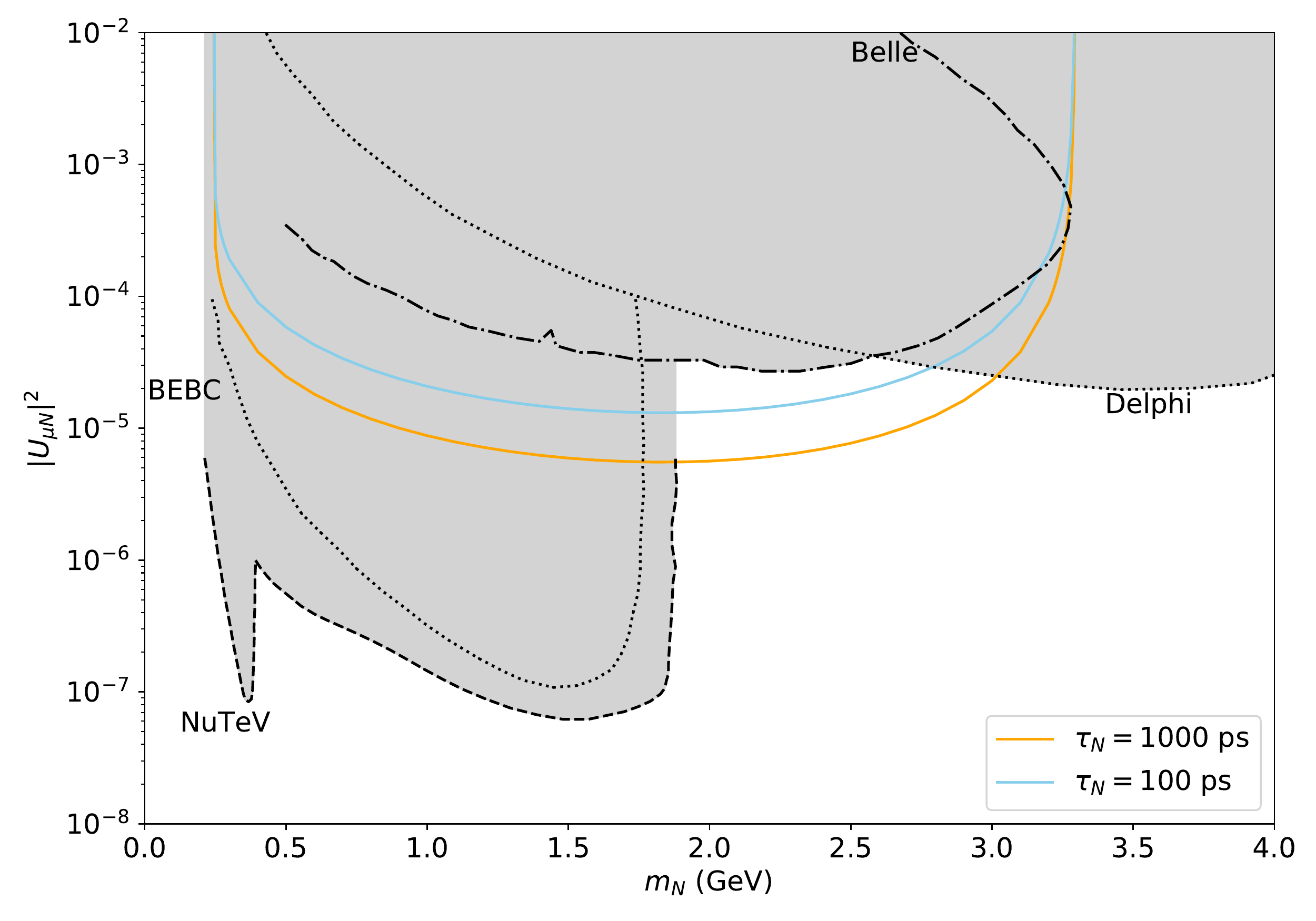}
\caption{The upper bound on the heavy-light mixing parameter $|U_{\mu N}|^2$ under the assumption that no positive signal about the processes is observed $B_s^0 \rightarrow D_s^- \mu^+ \mu^+ \pi^-$ on LHCb. The shaded regions represents the excluded region given by previous experiments including NuTeV~\cite{Vaitaitis:1999wq}, BEBC~\cite{CooperSarkar:1985nh}, Belle~\cite{Liventsev:2013zz} and Delphi~\cite{Abreu:1996pa}.}
\label{upper bound}
\end{figure}

On the other hand, suppose that such modes were not observed on the LHCb upgrade II, we can constrain the upper bound on the heavy-light mixing parameter $|U_{\mu N}|^2$ by requiring the total number of events to be lower than some threshold. Reference~\cite{Feldman:1997qc} shows that the experimental sensitivity on $|U_{\mu N}|^2$ at $95\%$ confidence level (C.L.) under a background-free environment is obtained for $\mathcal{N}_{\text{events}} = 3.09$. Thus, the upper bound on $|U_{\mu N}|^2$ at $95\%$ C.L. is obtained by requiring that 
\begin{equation}
    N_{B_s}\times \epsilon~(B_s \rightarrow D_s \mu \mu \pi) \times \mathcal{P}_N \times \mathcal{B}r_{\mathrm{avr}} = 3.09.
\label{bound}
\end{equation}
The numerical result of Equation~(\ref{bound}) is shown in Fig.~\ref{upper bound}. For comparison, we also include currently known upper bounds on $|U_{\mu N}|^2$ given by other experiments including NuTeV~\cite{Vaitaitis:1999wq}, BEBC~\cite{CooperSarkar:1985nh}, Belle~\cite{Liventsev:2013zz}, and Delphi~\cite{Abreu:1996pa} in the plot. The plot shows that during the  mass region $[1~\mathrm{GeV}, 3~\mathrm{GeV}]$, the channel gives a comparable or even stronger constraint on the size of $|U_{\mu N}|^2$. Specifically, for $m_N$ equals 2~GeV, the constraint on $|U_{\mu N}|^2$ goes as low as $6\times 10^{-6}$.

\section{Summary and conclusions}
\label{section four}
In this paper, we study the lepton-number-violating~(LNV) process $B_s^0 \rightarrow D_s^- \mu^+ \mu^+ \pi^-$ and its \textit{CP}-conjugate process $\bar{B_s^0} \rightarrow D_s^+ \mu^- \mu^- \pi^+$ that are induced by two nearly-degenerate Majorana neutrinos and explore the possibility for searching for the \textit{CP} violation in such processes. We point out that the physics origin of the \textit{CP} violation is  $\vartheta_{12}$, which is defined as the difference between the \textit{CP}-odd phases of mixing parameters between two generations of heavy neutrinos with the normal ones. The \textit{CP} violation becomes considerable when the masses of the two generations of heavy neutrinos are nearly degenerate but have a nonzero difference $\Delta m_{N}$. The numerical result draws the following conclusion. First, the relative size of the \textit{CP} violation $\mathcal{A}_{CP}$ is only the function of the mass difference $\Delta m_{N}$ and $\vartheta_{12}$ and $\mathcal{A}_{CP}$ is nearly independent of the absolute mass of the lighter heavy neutrino in the mass region we considered. Second, $\mathcal{A}_{CP}$ reaches its maximum when $\Delta m_N$ is around the size of the decay width of the intermediate sterile neutrino and the maximum value depends on the \textit{CP}-odd phase difference $\vartheta_{12}$. It should be noted that the above conclusion is drawn under the assumption that the two Majorana neutrinos have approximately the same mixing parameters with the three normal neutrinos.

We also analyze the possibility that such \textit{CP} violation is observed by LHCb during its upgrade II. It is shown that under a current constraint on the heavy-light neutrino mixing parameter, namely, $10^{-4} < |U_{\mu N}|^2 < 10^{-5}$, it is possible that such a \textit{CP} violation can be observed with the LHCb experimental ability. We also give the upper bound on the heavy-light mixing parameter under the assumption that no positive signal of the processes is observed.  The result shows that such modes can give a complementary constraint on the heavy-light mixing parameter compared with previous experiments including NuTeV, BEBC, Belle, and Delphi in the mass region 1~GeV $< m_N <$ 3~GeV. Thus, we note that it is worthwhile searching for such modes on LHCb due to the possibility of both observing new types of \textit{CP} violation in $B_s$ meson decays and setting complementary constraints on $|U_{\mu N}|^2$.

\section{Acknowledgement}
This work is supported by National Natural Science Foundation of China (Grant No.~12075003).

\section*{APPENDIX}
\appendix
\section{THE FOUR-BODY PHASE SPACE $d \Phi_4$}
\label{appendix a}
\begin{figure}[htb] 
\includegraphics[width=8.8cm]{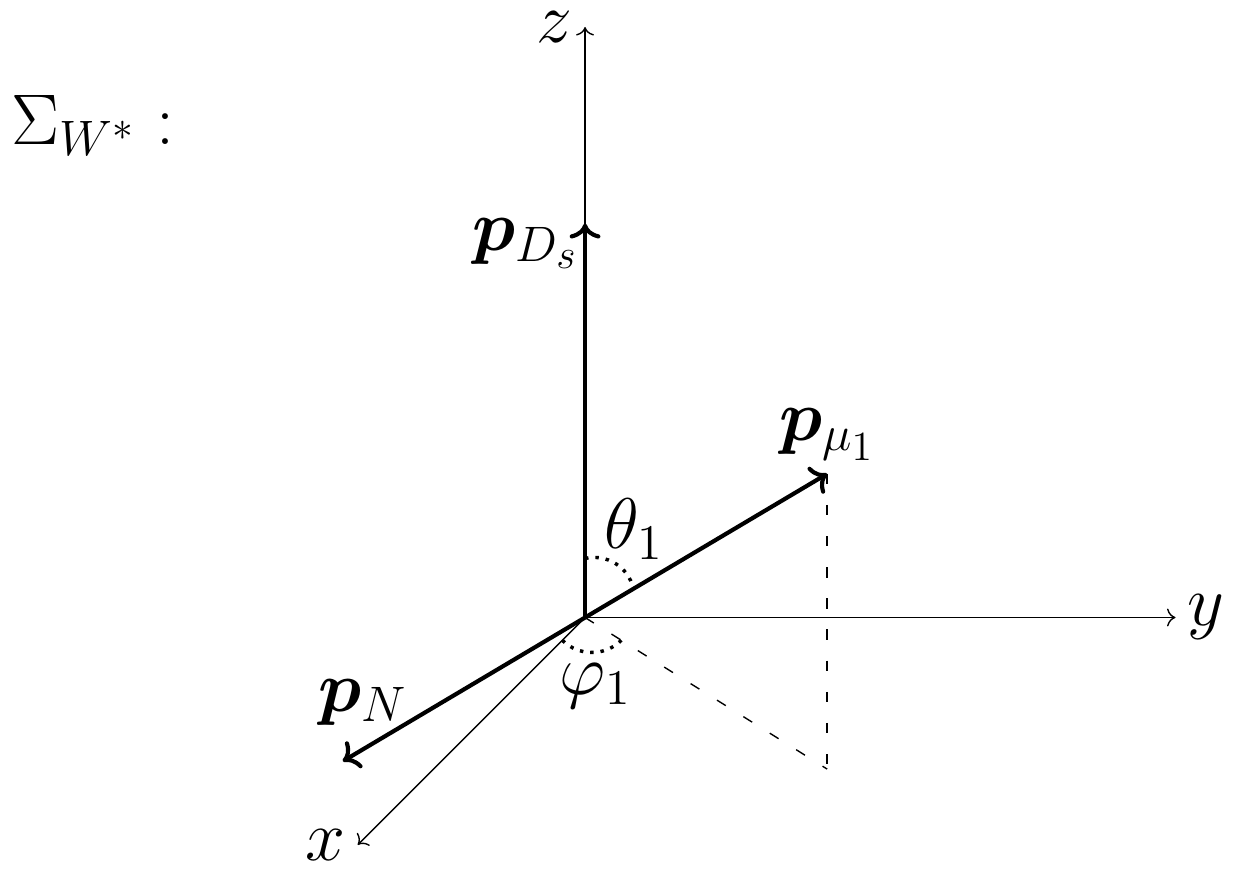} 
\includegraphics[width=8.8cm]{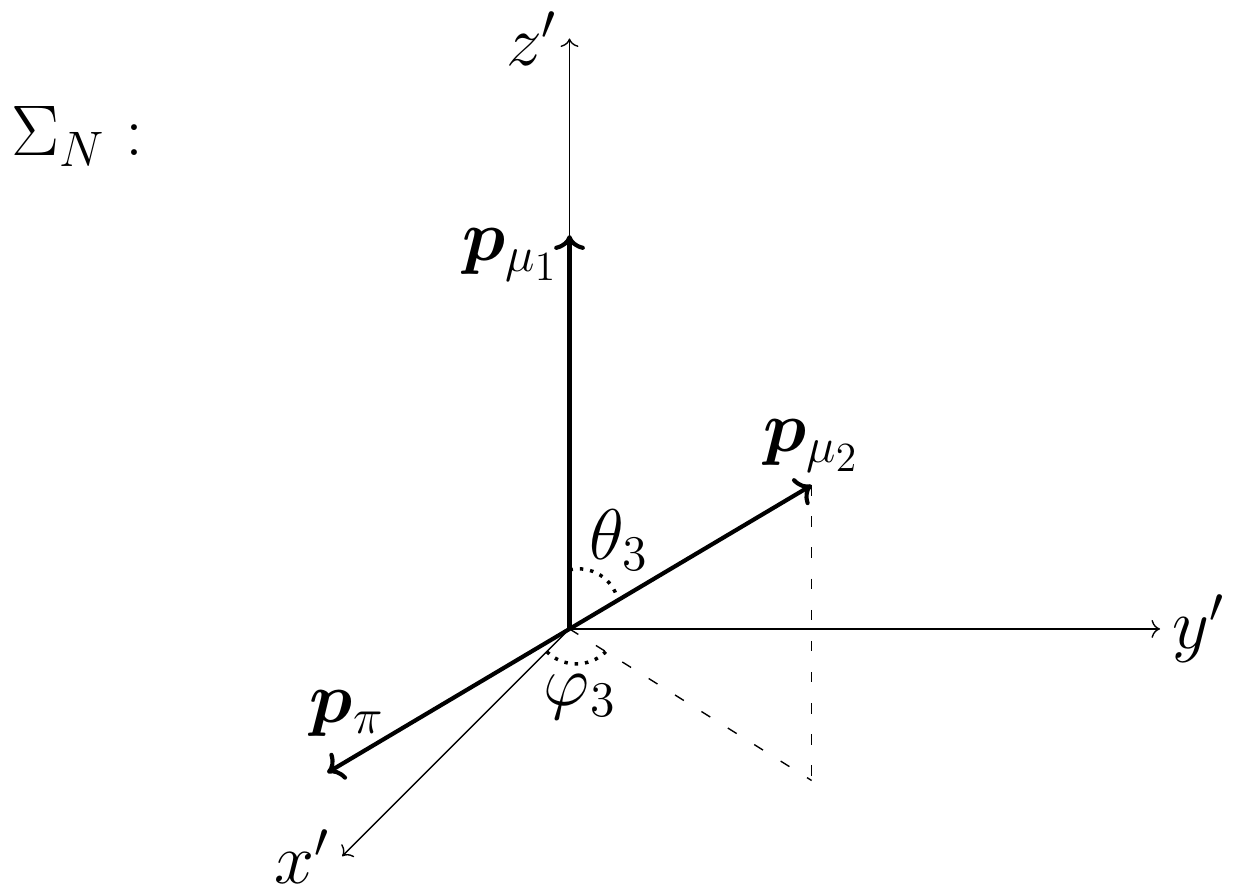}
\caption{The reference frames we use and the definitions of the solid angles.}
\label{reference frame}
\end{figure}

The four-body phase space $d\Phi_4$ is
\begin{equation}
    d\Phi_4 = \frac{1}{2 m_{B_{s}}}\frac{1}{(2\pi)^8} d_4,
\end{equation}
where the factor $d_4$ is defined as
\begin{equation}
    d_4 = \frac{\ud^3 \bm{p}_{D_s}}{(2 E_{D_s})}\frac{d^3 \bm{p}_{1}}{(2 E_{1})}\frac{d^3 \bm{p}_{2}}{(2 E_{2})}\frac{d^3 \bm{p}_{\pi}}{(2 E_{\pi})}\delta^4(p_{B_s} - p_{D_s} - p_{\mu_1} - p_{\mu_2} - p_{\pi}).
\end{equation} 
Remember the fact that
\begin{equation}
    \frac{d^3 \bm{p}_{N}}{(2 E_{N})} = d^4 p_N \delta_+(p_N^2 - m_N^2),
\end{equation}
then it is straightforward to prove that $d_4$ can be factorized as
\begin{equation}
    d_4 = d_3(B_s \rightarrow D_s \mu_1 N) d_2(N \rightarrow \mu_2 \pi)dp_N^2.
\end{equation}
Here, $d_3(B_s \rightarrow D_s \mu_1 N)$ is the corresponding factor in the three-body phase space for the subprocess $B_s \rightarrow D_s \mu_1 N$ and $d_2(N \rightarrow \mu_2 \pi)$ is that for $N \rightarrow \mu_2 \pi$. $p_N$ is the four-momenta of the intermediate sterile neutrino. In the following we set $t_2 = p_N^2$. In the rest frame of $\mu_1-N$, $d_3(B_s \rightarrow D_s \mu_1 N)$ can be simplified as
\begin{equation}
    d_3(B_s \rightarrow D_s \mu_1 N) = \frac{1}{64 m_{B_s}^2}\lambda^{\frac{1}{2}}(m_{B_s}^2, m_{D_s}^2, t_1)\lambda^{\frac{1}{2}}(t_1, m_{\mu}^2, m_N^2)\frac{1}{t_1} d t_1 d \Omega^{*}_{\mu_1}d\Omega_{D_s},
\end{equation}
where $t_1 = (p_N + p_{\mu_1})^2 = (p_{B_s} - p_{D_s})^2$, and $d\Omega^*_{\mu_1} = d\cos\theta_1d\varphi_1$ is the solid angle of $\mu_1$ in the $\mu_1 - N$ rest frame, while $d\Omega_{D_s} = d\cos \theta_2 \varphi_2$ is that of $D_s$ in the rest frame of $B_s$. The function $\lambda(x, y, z)$ is the kinematic K{\"a}llen function, $\lambda(x, y, z) = x^2 + y^2 + z^2 - 2xy -2yz- 2xz$. In the rest frame of $N$, $d_2(N \rightarrow \mu_2 \pi)$ is written as
\begin{equation}
    d_2(N \rightarrow \mu_2 \pi) = \frac{1}{8 t_2}\lambda^{\frac{1}{2}}(t_2, m_{\mu}^2, m_{\pi}^2) d\Omega_{\mu_2},
\end{equation}
where $d\Omega_{\mu_2} = d\cos\theta_3 d\varphi_3$ is the solid angle of $\mu_2$ in the rest frame of $N$. As is shown in the next part, the square of the amplitude $|\mathcal{M}|^2$ is independent of $\varphi_1$ and $\Omega_{D_s}$; thus, they can be integrated and give a factor $8\pi^2$. As a result, the four-body phase space $d_4$ is
\begin{equation}
    d_4 = \frac{\pi^2}{64 m_{B_s}^2}\lambda^{\frac{1}{2}}(m_{B_s}^2, m_{D_s}^2, t_1)\lambda^{\frac{1}{2}}(t_1, m_{\mu}^2, m_N^2)\lambda^{\frac{1}{2}}(t_2, m_{\mu}^2, m_{\pi}^2)\frac{1}{t_1 t_2}d t_1 d t_2 d \cos \theta_1 d\cos\theta_3 d\varphi_3.
\end{equation}
Here, $t_1$ and $t_2$ satisfy that $(\sqrt{t_2} + m_{\mu})^2 \leq t_1 \leq (m_{B_s} - m_{D_s})^2$ and $t_2 \geq (m_{\mu} + m_{\pi})^2$. The definitions of $\theta_1, \phi_1, \theta_3, \phi_3$  and the reference frames we used are shown in Fig.~\ref{reference frame}. 

\section{EXPLICIT FORM OF THE SQUARED MATRIX $M$}
\label{appendix b}
For simplicity, we define in the following that
\begin{equation}
    K = G_F^2 V_{cb}V_{ud}f_{\pi}
\end{equation}
and
\begin{equation}
    T^{\pm} = \bar{u}(\mu_2)\slashed{p}_{\pi}\gamma_{\mu}(1 \mp \gamma_5)v(\mu_1)\langle D_s^- |c(0) \gamma_{\mu} b(0)| B_s^0 \rangle.
\end{equation}
$T^{\pm}$ contains all the spinor structures of the decaying processes. It can be checked that $|T^{+}|^2 = |T^{-}|^2 \equiv |T|^2$. Then the elements of the squared amplitude matrix $M$ are written as
\begin{equation}
    M_{ij} = |K|^2 m_{N_i}m_{N_j}P_{N_i}P_{N_j}^*|T|^2
\label{M matrix}
\end{equation}
Apart from $P_{N_i}P_{N_j}^*$, other factors in $M_{ij}$ are always real. Thus, the real/imaginary part of $M_{12}$ is proportional to the real/imaginary part of $P_{N_1}P_{N_2}^*$. The completed form of $|T|^2$ is
\begin{equation}
    |T|^2 = C_{00} f_{0}^2(t_1) + C_{01} f_{0}(t_1)f_{+}(t_1) + C_{11} f_{+}^2(t_1).
\end{equation}
Here, the coefficients are
\begin{equation}
\begin{aligned}
C_{00} &= \frac{8}{t_1^2}(m_{B_s}^2 - m_{D_s}^2)^2 \Big\{m_{\pi}^2[t_1(p_{\mu_1} \cdot p_{\mu_2}) - 2(q \cdot p_{\mu_1})(q \cdot p_{\mu_2})] + 4(p_{\pi} \cdot p_{\mu_2})(q \cdot p_{\pi})(q \cdot p_{\mu_1}) - 2 t_1 (p_{\pi} \cdot p_{\mu_1})(p_{\pi} \cdot p_{\mu_2})  \Big\},\\
C_{01} & = \frac{16}{t}(m_{B_s}^2 - m_{D_s}^2)\Big\{ (q \cdot p_{\mu_1}) \big[2(p_{\pi}\cdot p_{\mu_2})(P \cdot p_{\pi}) - m_{\pi}^2(P\cdot p_{\mu_2})\big] + (P \cdot p_{\mu_1})\big[2(p_{\pi}\cdot p_{\mu_2})(q \cdot p_{\pi}) - m_{\pi}^2(q \cdot p_{\mu_2}) \big] \Big\}, \\
C_{11} & = 8\Big\{P^2[m_{\pi}^2(p_{\mu_1} \cdot p_{\mu_2}) - 2 (p_{\pi} \cdot p_{\mu_1})(p_{\pi} \cdot p_{\mu_2})] - 2 m_{\pi}^2(P \cdot p_{\mu_1})(P \cdot p_{\mu_2}) + 4 (p_{\pi} \cdot p_{\mu_2})(P \cdot p_{\pi})(P \cdot p_{\mu_1})\Big\},
\end{aligned}
\label{coefficents}
\end{equation}
where we define that
\begin{equation}
    P = p_{B_s} + p_{D_s} - \frac{m_{B_s}^2 - m_{D_s}^2}{t_1}q.
\end{equation}
Most inner products in Equation~(\ref{coefficents}) can be written as functions of $t_1,t_2,\theta_1,\theta_3,\varphi_3$ directly, namely,
\begin{equation}
\begin{aligned}
    &p_{D_s}\cdot q = \frac{1}{2}(m_{B_s}^2 - m_{D_s}^2 - t_1),~ p_{D_s}\cdot p_{\mu_1} = E_{D_s}^*E_{\mu_1}^* - |\bm{p}_{D_s}^*||\bm{p}_{\mu_1}^*|\cos\theta_1,~q\cdot p_{\mu_1} = \frac{1}{2}(t_1 + m_{\mu}^2 - t_2),\\
    &p_{\mu_2}\cdot p_{\pi} = \frac{1}{2}(t_2 - m_{\mu}^2 - m_{\pi}^2),~~~ p_{\mu_1}\cdot p_{\pi} = E_{\mu_1}E_{\pi} + |\bm{p}_{\mu_1}||\bm{p}_{\mu_2}|\cos \theta_3, ~~~~p_{\mu_1}\cdot p_{\mu_2} = E_{\mu_1}E_{\mu_2} - |\bm{p}_{\mu_1}||\bm{p}_{\mu_2}|\cos \theta_3.
\end{aligned}
\end{equation}
Here, we list the explicit forms for the energies and three-momenta that appear in the above terms (the superscript ``$*$" represents the value in the $W^*$ rest frame),
\begin{equation}
\begin{aligned}
    &E_{D_s}^* = \frac{1}{2\sqrt{t_1}}(m_{B_s}^2 - m_{D_s}^2 - t_1), ~E_{\mu_1}^* = \frac{1}{2\sqrt{t_1}}(t_1 + m_{\mu}^2 - t_2),~ |\bm{p}_{D_s}^*| = \frac{1}{2\sqrt{t_1}}\lambda^{\frac{1}{2}}(m_{B_s}^2, m_{D_s}^2, t_1),\\
    &|\bm{p}_{\mu_1}^*| = \frac{1}{2\sqrt{t_1}}\lambda^{\frac{1}{2}}(t_1, m_{\mu}^2, t_2),\qquad E_{\mu_1} = \frac{1}{2\sqrt{t_2}}(t_1 - m_{\mu}^2 - t_2), ~~ E_{\mu_2} = \frac{1}{2\sqrt{t_2}}(t_2 - m_{\mu}^2 + m_{\pi}^2),\\
    &E_{\pi} = \frac{1}{2\sqrt{t_2}}(t_2 - m_{\mu}^2 + m_{\pi}^2), \qquad |\bm{p}_{\mu_1}| = \frac{1}{2\sqrt{t_2}}\lambda^{\frac{1}{2}}(t_1, m_{\mu}^2, t_2), ~~|\bm{p}_{\mu_2}| = \frac{1}{2\sqrt{t_2}}\lambda^{\frac{1}{2}}(t_2, m_{\mu}^2, m_{\pi}^2),\\
    &E_N^* = \frac{1}{2\sqrt{t_1}}(t_1 + t_2 - m_{\mu}^2), \qquad ~  |\bm{p}_N^*| = \frac{1}{2\sqrt{t_1}}\lambda^{\frac{1}{2}}(t_1, t_2, m_{\mu}^2).
\end{aligned}
\label{energy}
\end{equation}
Note that $q = p_{\mu_1} + p_{\mu_2} + p_{\pi}$, so $q \cdot p_{\mu_2}$ and $q \cdot p_{\pi}$ can also be written directly. In order to get $p_{D_s}\cdot p_{\mu_2}$ and $p_{D_s}\cdot p_{\pi}$, we need to do a Lorentz transformation from the $W*$ rest frame to the $N$ rest frame on $p_{D_s}$, since only the four-momenta vector of $D_s$ in the $W*$ rest frame can be written directly. We need to first rotate the $z$ axis of $\Sigma_{W^*}$ to the direction of $\bm{p}_{\mu_1}$, then boost the four-momenta vector of $p_{D_s}$ from the $W^*$ rest frame to the $N$ rest from. In the $W^*$ rest frame where the $z$ axis points to the direction of $\bm{p}_{\mu_1}$, the four-momenta vector of $D_s$ is
\begin{equation}
    p_{D_s}^* = (E_{D_s}^*, ~|\bm{p}^*_{D_s}|\sin\theta_1, ~0, ~|\bm{p}^*_{D_s}|\cos\theta_1). 
\end{equation}
The Lorentz boost from the $W^*$ rest frame to the $N$ rest frame is
\begin{equation}
    B(\bm{v}_N)
    = \left(\begin{array}{cccc}
    \gamma_N & ~0 & ~0 & ~\gamma_N|\bm{v}_N|\\
    0 & ~1 & ~0 & ~0\\
    0 & ~0 & ~1 & ~0\\
    \gamma_N|\bm{v}_N| & ~0 & ~0 & ~\gamma_N
    \end{array}\right),
\end{equation}
where $\gamma_N$ and $\bm{v}_N$ are the Lorentz factor and the speed of $N$ in the $W^*$ rest frame, respectively,
\begin{equation}
    \gamma_N = \frac{E_N^*}{m_N}, ~~~|\bm{v}_N| = \frac{|\bm{p}_N^*|}{E_N^*}.
\end{equation}
Here, $E_N^*$ and $|\bm{p}_N^*|$ are given in Equation~(\ref{energy}). As a result, $p_{D_s}\cdot p_{\mu_2}$ and $p_{D_s}\cdot p_{\pi}$ are written as
\begin{equation}
\begin{aligned}
    &p_{D_s}\cdot p_{\mu_2} = \gamma_N(E_{D_s}^* + |\bm{v}_N||\bm{p}_{D_s}^*|\cos \theta_1)E_{\mu_2} - |\bm{p}_{D_s}^*||\bm{p}_{\mu_2}|\sin \theta_1 \sin \theta_3\cos \varphi_3 - \gamma_N(|\bm{p}_{D_s}^*|\cos \theta_1 + |\bm{v}_N|E_{D_s}^*)|\bm{p}_{\mu_2}|\cos\theta_3, \\
    &p_{D_s} \cdot p_{\pi} ~= \gamma_N(E_{D_s}^* + |\bm{v}_N||\bm{p}_{D_s}^*|\cos \theta_1)E_{\pi} + |\bm{p}_{D_s}^*||\bm{p}_{\mu_2}|\sin \theta_1 \sin \theta_3\cos \varphi_3 + \gamma_N(|\bm{p}_{D_s}^*|\cos \theta_1 + |\bm{v}_N|E_{D_s}^*)|\bm{p}_{\mu_2}|\cos\theta_3
\end{aligned}
\end{equation}

\section{DECAY WIDTH FOR $B_s \rightarrow D_s \mu_1 N_i$ AND FOR $N_i \rightarrow \mu_2 \pi$}
\label{appendix c}

\begin{figure}[htb]
\includegraphics[width = 8cm]{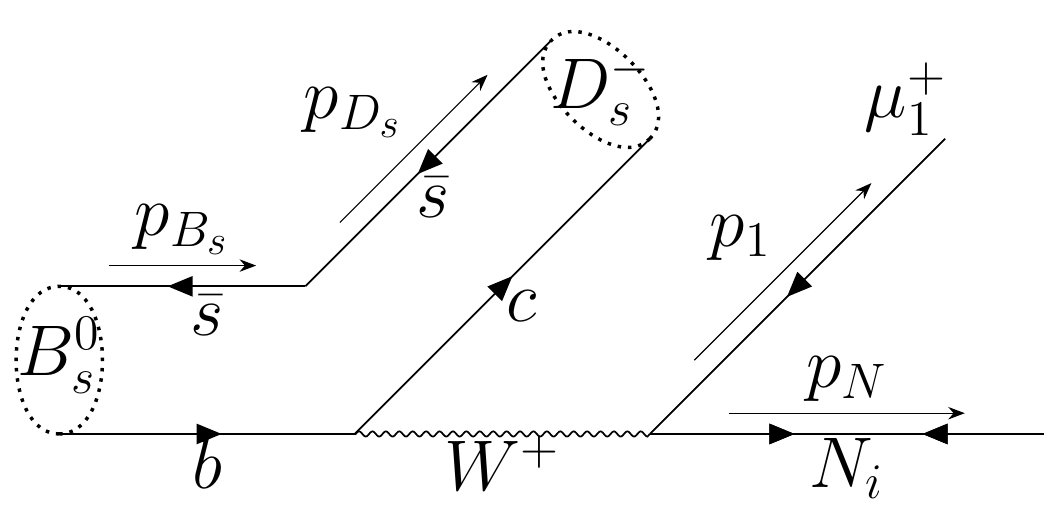}
\caption{Feynman diagram for $B_s \rightarrow D_s \mu_1 N_i$}
\label{feynman diagram2}
\end{figure}

The Feynman diagram for $B_s \rightarrow D_s \mu_1 N_i$  is shown in Fig.~\ref{feynman diagram2}. The amplitude for the process is
\begin{equation}
    i\mathcal{M}(B_s \rightarrow D_s \mu_1 N_i) = \frac{G_F}{\sqrt{2}}U_{\mu N_i}V_{cb}\bar{u}(p_{N_i})\gamma_{\mu}(1 - \gamma_5)v(p_1)\langle D_s^-|c(0)\gamma^{\mu}b(0)|B_s^0 \rangle,
\end{equation}
Integrating the amplitude over three-body phase space, we have the decay width for $B_s \rightarrow D_s \mu_1 N_i$,
\begin{equation}
    \Gamma(B_s \rightarrow D_s \mu_1 N_i) = \frac{G_F^2 V_{cb}^2}{384\pi^3}\frac{1}{m_{B_s}^3}\int_{t_{min}}^{t_{max}} dt \frac{1}{t^2} \lambda^{\frac{1}{2}}(m_{B_s}^2, m_{D_s}^2, t)\lambda^{\frac{1}{2}}(m_{\mu}^2, m_{N_i}^2, t)\Big[f_+^2(t)D_1(t) + f_0^2(t)D_0(t)\Big],
\end{equation}
where $t_{min} = (m_{N_i} + m_{\mu})^2$, $t_{max} = (m_{B_s} - m_{D_s})^2$ and coefficients $D_1(t)$ and $D_0(t)$ are
\begin{equation}
\begin{aligned}
    D_1(t) &= \Big[(t - m_{D_s}^2)^2 - 2m_{B_s}^2(t + m_{D_s}^2) + m_{B_s}^4 \Big]\Big[2t^2 - t m_{N_i}^2 + m_{\mu}^2(2m_{N_i}^2 - t) - m_{N_i}^2 - m_{\mu}^4 \Big]\\
    D_0(t) &= 3(m_{B_s}^2 - m_{D_s}^2) \Big[t m_{N_i}^2 + m_{\mu}^2(2 m_{N_i}^2 + t) - m_{N_i}^4 - m_{\mu}^4\Big].
\end{aligned}
\end{equation}
As for the process $N_i \rightarrow \mu_2 \pi$, the decay width is well known in the literature,
\begin{equation}
    \Gamma(N_i \rightarrow \mu \pi) =\frac{G_F^2}{16\pi}|V_{ud}|^2|U_{\mu N_i}|^2 f_{\pi}^2 m_{N_i}\lambda^{1/2}(m_{N_i}^2, m_{\mu}^2, m_{\pi}^2)\left[\left(1 - \frac{m_{\mu}^2}{m_{N_i}^2}\right)^2 - \frac{m_{\pi}^2}{m_{N_i}^2}\left(1 + \frac{m_{\mu}^2}{m_{N_i}^2}\right)\right] .
\end{equation}

\section{TOTAL DECAY WIDTH FOR THE INTERMEDIATE MAJORANA NEUTRINO}
\label{appendix d}
\begin{figure}[htb]
\includegraphics[width = 8 cm]{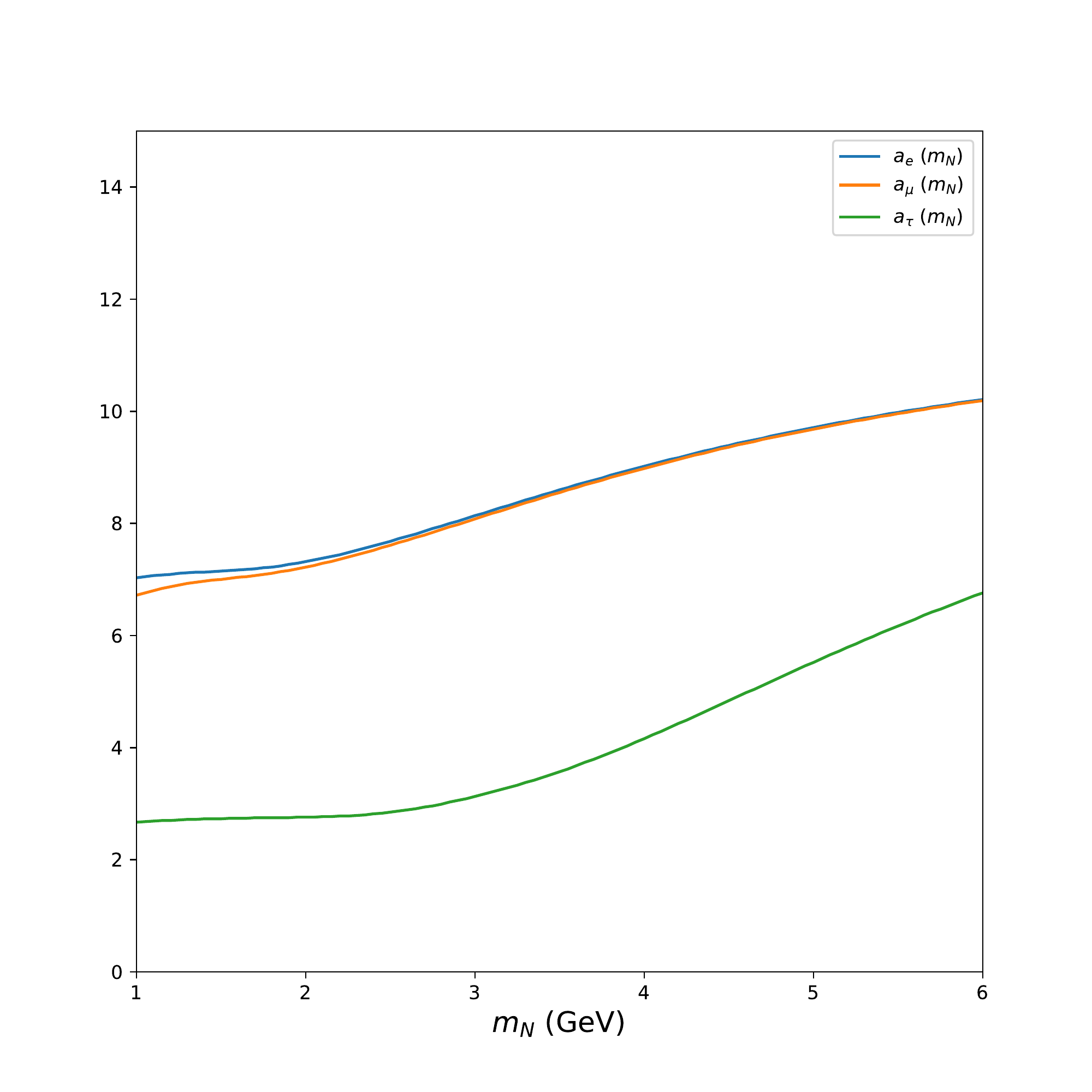}
\caption{Numerical values of $a_l(m_N)$}
\label{decay width factor}
\end{figure}

Although $\Gamma_{N_i}$ can be calculated through the channel-by-channel approach, which sums up the partial decay rates for all the leptonic and semileptonic decay modes of $N$~\cite{Atre:2009rg}, for neutrino mass larger than 1~GeV the uncertainties of the hadronic parameters such as the decay constants of the final hadronic states are large. Since we are interested in the mass range between 1 and 4~GeV, here we use the inclusive approach introduced in Ref.~\cite{Helo:2010cw}, which approximates the semileptonic decays of $N$ by its decays into free quark-antiquark pairs and leptons and the approximation is better than the channel-by-channel method for neutrino mass more than 1~GeV~\cite{Bondarenko:2018ptm}. We refer to Ref.~\cite{Helo:2010cw} for details of the calculation. Then all the decay channels of $N$ are three-body decays and the partial decay rates are proportional to $m_N^5$. Thus, the total decay width can be written as
\begin{equation}
    \Gamma_{N_i} = \frac{G_F^2 m_{N_i}^5}{96\pi^3}\sum_{l=e, \mu, \tau}|U_{lN}|^2 a_{l}(m_{N_i}),
\label{total decay width}
\end{equation}
where $a_{l}(m_{N_i})$ are dimensionless functions of $m_{N_i}$, and the numerical values of $a_{l}(m_{N_i})$ are presented in Fig.~\ref{decay width factor}.

\end{document}